# Transferred plasma jet from a dielectric barrier discharge for processing of poly(dimethylsiloxane) surfaces

Fellype do Nascimento[1*], Munemasa Machida[2], Mara A. Canesqui[1], and Stanislav Moshkalev[1]

[1]*Center for Semiconductor Components and Nanotechnologies – State University of Campinas, C. P. 6101, Campinas, CEP 13083-870, Brazil*

[2] *Institute of Physics "Gleb Wataghin" – State University of Campinas, Campinas, CEP 13083-859, Brazil*

## Abstract

In this work we studied processing of poly(dimethylsiloxane) (PDMS) surfaces using dielectric barrier discharge (DBD) plasma in two different assemblies, one using the primary plasma jet obtained from a conventional DBD and the other using a DBD plasma jet transfer. The evolution of water contact angle (WCA) in function of plasma processing time and in function of aging time as well as the changes in the surface roughness of PDMS samples for both plasma treatments have been studied. We also compared vibrational and rotational temperatures for both plasmas and for the first time the vibrational temperature ($T_{vib}$) for the transferred plasma jet has been shown to be higher as compared with the primary jet. The increment in the $T_{vib}$ value seems to be the main reason for the improvements in adhesion properties and surface wettability for the transferred plasma jet. Possible explanations for the increase in the vibrational temperature are presented.

## 1. Introduction

Atmospheric pressure plasmas have received considerable attention in recent years due to their versatility, easiness of operation and low cost compared to plasmas in vacuum environment [1,2]. Cold atmospheric pressure plasma jets are characterized by low temperature that is especially important for modification or activation of surfaces of soft materials like polymers or biological tissues, without damaging them [3-6]. Dielectric barrier discharge (DBD) plasma is a kind of atmospheric pressure plasma in which the discharges are produced between two electrodes with at least one of them covered with a dielectric material (glass or ceramic in most cases) [7,8].

* E-mail: fellype@gmail.com

In microfluidics and biomedical applications the flexible poly(dimethylsiloxane) (PDMS) polymer is widely employed due to its chemical inertness, thermal stability and low cost. However, its low surface energy limits its adhesion properties, and surface activation by plasma treatments or chemical processing is frequently used in particular to improve its hydrophilicity [9-13].

The mechanisms concerning the improvement of adhesion to polymer surfaces by plasma processing are currently under study and may involve surface cleaning and activation, surface functionalization and modification of surface morphology [14-16]. There is a correlation between the water contact angle (WCA) and adhesion properties, with lower contact angles resulting in better adhesion, that is an indicative of the better surface wettability, as observed in a number of works [17-19]. Modifications in the polymer surface roughness that can be associated with better adhesion were also observed for various polymeric materials treated with plasmas [15,20-22]. For the case of PDMS, earlier works reported changes in the surface morphology and/or creation of functional groups in the surface after treatments using atmospheric pressure plasmas with different gases [10,12,13,23-25]. The plasma jet was reported to change the surface chemistry destructing methyl groups (Si–$CH_3$) and introducing hydrophilic silanol groups (Si–OH) and also to reduce the surface roughness increasing its contact area [16,24,26].

Studies of transfer of plasma jets using atmospheric pressure plasmas started with Lu *et al* [27] when he discovered that a plasma plume was able to ignite a secondary plasma discharge inside a dielectric tube with a He gas flow passing through it. More recently, Kostov *et al* [28] started to study the transfer of plasma jets using a conducting wire inside a long plastic tube and demonstrated the possibility of generating a plasma jet remotely at the output of a plastic tube with no ignition of plasma inside it. Thereafter Kostov *et al* developed an application of this technique for microbial decontamination [29], and in a recently conducted study Xia *et al* [30] have shown that the velocity of plasma bullets from a transferred plasma jet increases with the increment of the length of the conducting wire and the plastic tube.

In the current study we analyzed the differences in the treatment of PDMS surfaces when a primary plasma jet or a transferred plasma jet is applied. The primary plasma jet is produced using a DBD device that was built using a 5C22 thyratron valve and a ferrite transformer [6,13,31]. We observed that the processes of reduction and recovery of WCA after the processing by these two types of treatment are much different. Using the transferred plasma jet to treat the surfaces, the WCA reduction is faster and its recovery is slower than using the primary plasma jet, resulting in better adhesion between two similar PDMS samples for the former. For the first time we report here that the vibrational temperature ($T_{vib}$) obtained in the transferred plasma jet is significantly higher than that obtained in the primary jet, and it seems to be the main reason for the improvements in adhesion properties and surface wettability. We propose also a correlation between the vibrational temperature of the plasma jets and the tensile strength supported in adhesion between two PDMS samples: the higher the $T_{vib}$ value, the higher the tensile strength supported. The changes in the vibrational temperature when switching from one assembly to other are probably related to the increase in the rate in which the electromagnetic field energy is transferred into the energy of vibrational states.

## 2. Vibrational excitation in $N_2$ plasmas

In DBD plasmas there are several processes that can change the number density of $N_2$ molecules in vibrationally excited states, increasing the vibrational energy of the gas: collisions with electrons, collisions with metastable atoms and ions, energy transfer by collisions between vibrationally excited molecules, and conversion of vibrational energy in translational and/or rotational motion of the molecules [32-35].

In the energy diagram shown in Fig. 1 we can see that metastable helium He(2 $^3$S) atoms (denoted hereafter as $He^M$) and argon ions $Ar^+(^3P_{3/2})$ have energy levels high enough to ionize $N_2$ molecules to states $N_2^+(A\ ^2\Pi_u$ and $B\ ^2\Sigma_u^+)$ for He, and $N_2^+(X\ ^2\Sigma_g^+)$ for $Ar^+$. The $N_2^+$ ions and argon metastable states $Ar(^3P_2$ and $^3P_0)$ (denoted hereafter as $Ar^M$) can further collide with $N_2$ molecules and produce excited states $N_2(C\ ^3\Pi_u$ and $B\ ^3\Pi_g)$ and metastable states $N_2(A\ ^3\Sigma_u^+$ and $a'\ ^1\Sigma_u^-)$ (denoted as $N_2^M$). Collisions between two metastable $N_2$ molecules and between a $N_2^+$ with ground state $N_2(X\ ^1\Sigma_g^+)$ molecule can also produce vibrationally excited $N_2$.

$N_2$ molecules in a ground state are vibrationally excited by electron impact at a rate of $4\times10^{-9}$ $cm^3\ s^{-1}$, for an electron temperature of 1 eV in DBD plasmas [36]. Additionally, an electronically excited $N_2$ molecule ($N_2^*$) produced by collisions with $N_2^+$, $N_2^M$, $Ar^+$, $Ar^M$, and/or $He^M$ may be found in a vibrationally excited states v', since metastables and ions are able to perform vibrational excitation on the molecules [37,38].

Comparing the energy levels of $N_2^+$ ions with those of $He^M$ and $Ar^+$ (see Fig. 1), we can infer that the helium gas is likely to be the best choice in order to produce a plasma with a high degree of vibrational excitation.

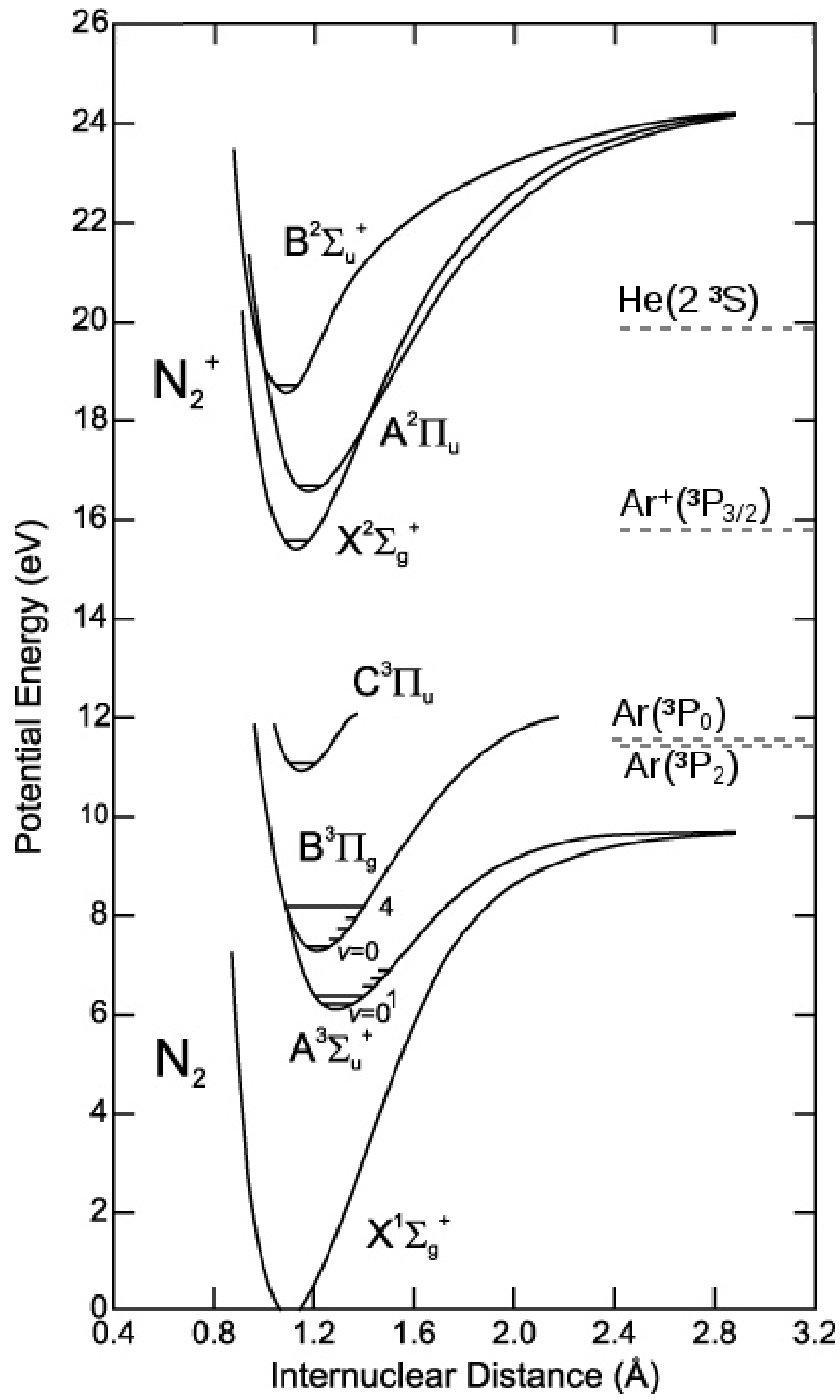

**Figure 1:** Energy level diagram for $N_2$ molecules, with references to the energies of He(2 $^3$S), Ar($^3P_2$, $^3P_0$), and $Ar^+$($^3P_{3/2}$) states (adapted from [39])

## 3. Experimental procedure

The experimental setups used to treat PDMS samples with the DBD plasmas are shown in Figure 2. Figure 2(a) shows the scheme for treatments using the primary plasma jet and Fig. 2(b) shows the scheme for the case when the transferred plasma jet is used.

In the plasma device, a continuous gas flow is injected inside the poly(vinyl chloride) (PVC) tube and high-voltage pulses are applied to the electrode inside the glass tube. A primary plasma discharge is formed in the region between the glass tube and the PVC tube producing a plasma jet leaving the tube exit of 10 mm in diameter, which is used for surface treatment of PDMS samples. For the transferred plasma jet configuration, an additional adapter to change the opening of the PVC tube is used to attach a flexible plastic tube, with a flexible conducting Cu wire inside it, to the opening of the PVC tube. The conducting wire is covered with plastic (polyethylene, 1.0 mm thick) and only the tips are exposed. The tip of the conducting wire does not touch the glass tube or any other parts of the DBD device, so that the primary plasma only touches. The lengths of the plastic tube and conducting wire used was 100 cm and 99.0 cm, respectively. There is no plasma formation along and inside the plastic tube. The plasma jet that leaves the plastic tube starts to form at the end of the conductor.

The voltage applied to the electrode was 15 kV in both assemblies when the He gas was used, and it was 30 kV when the $N_2$ gas was used (at lower voltages no stable discharges were produced for the $N_2$). The gas flow rate used to treat PDMS samples was the same for both assemblies and for all gases used, fixed at 4 L/min. The PDMS used was a commercial type (BISCO HT-6240, 250 µm thick). This PDMS does not require a surface cleaning or any other pre-treatments to be used – we just remove the protective layer and perform the plasma treatment. The dimensions of the samples were 15 x 15 $mm^2$.

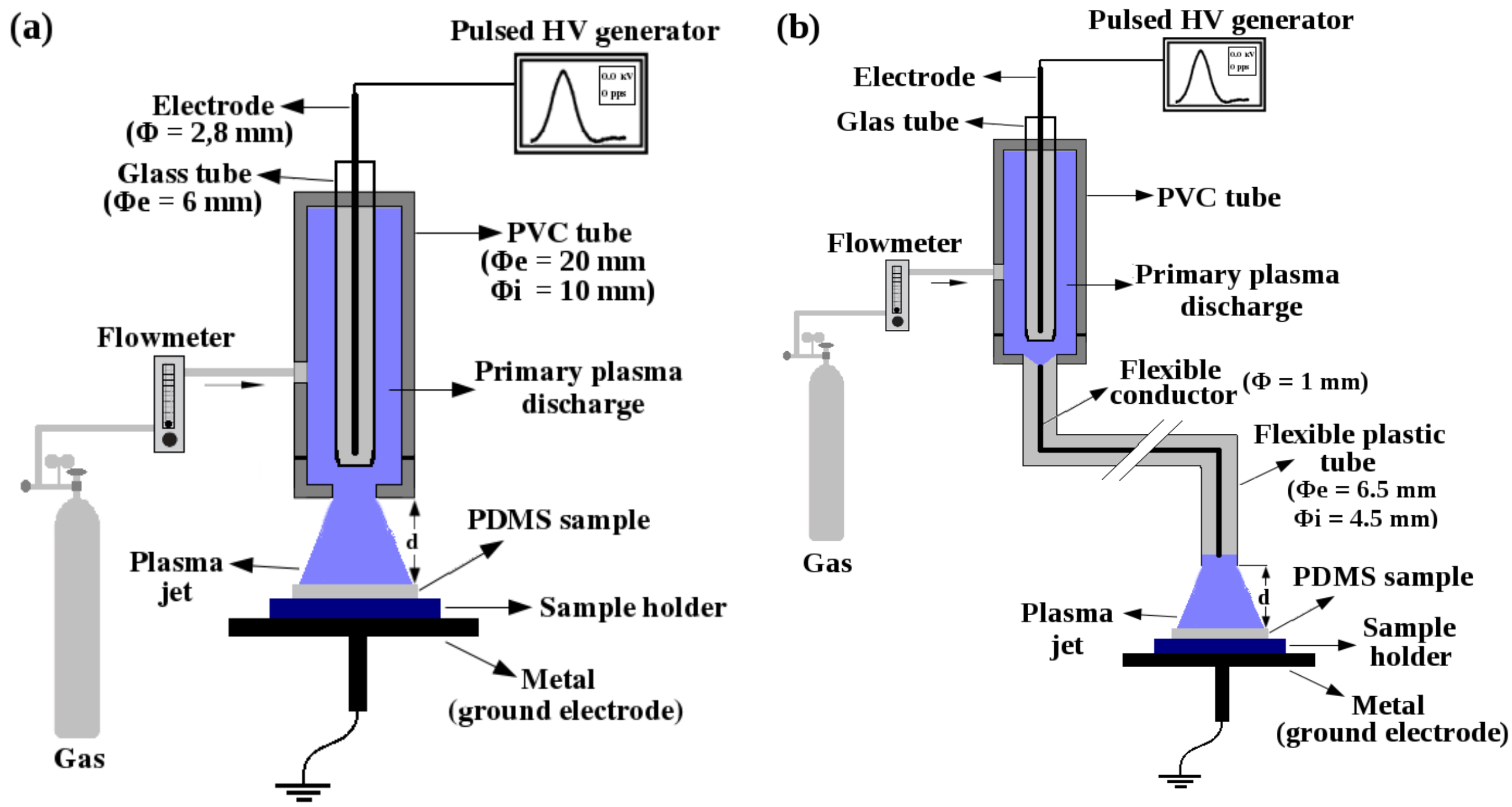

**Figure 2:** DBD schemes for plasma treatment of PDMS samples using: (a) plasma jet from primary discharge and (b) transferred plasma jet. Φe and Φi refer to external and internal diameters, respectively. The elements are out of scale.

One of the most noticeable difference between the two assemblies is in the diameter of the plasma jet and thus the area that can be treated with each plasma jet. In both cases the plasma jet spreads over the surface when in contact with it. However, due to the greater diameter, the primary plasma jet can treat a surface area larger than that can be treated using the transferred plasma jet. Furthermore, when performing the surface treatment of PDMS, the samples were placed at a distance of **d**=3 mm from the end of the PVC tube when the primary plasma jet was used, and the same distance **d** from the end of the plastic tube was adopted when the transferred plasma jet was used. In both cases we observed that the plasma jets were able to cover the entire samples, but the transferred plasma jet presented a coverage of the surface not as homogeneous as the primary plasma jet offered. The surface treatment made using $N_2$ gas provides a coverage similar to that with He gas.

For the WCA measurements to evaluate the PDMS surface wettability, we used a commercial photo camera and the ImageJ2 software [40].

In order to measure the tensile strength supported in adhesion between two similar PDMS samples, both samples were exposed to plasma treatment for the same time interval. Then, after the treatment the two parts were pressed against each other immediately and allowed to cure for two days at room temperature. After this period, tests were conducted to determine the tensile strength supported by the adhesion between samples. The setup adopted for the adhesion tests is shown in Fig. 3. The back sides of samples (that were not treated by plasma) were pasted with an epoxy-based commercial glue to the heads of stainless steel screws of 15mm in diameter and an increasing tensile force was applied until the adhesion failure (samples detachment) [13].

The vibrational temperatures were determined using the SpecAir software [41] and measurements of spectral emissions were performed using an Andor 303i spectrometer equipped with an iStar DH720 iCCD detector and a 1200 lines/mm grating. A 150 lines/mm grating was also used in order to get an overview of the entire spectra. The light emitted by the plasmas was collected with a lens and transported to the spectrometer through an optical fiber.

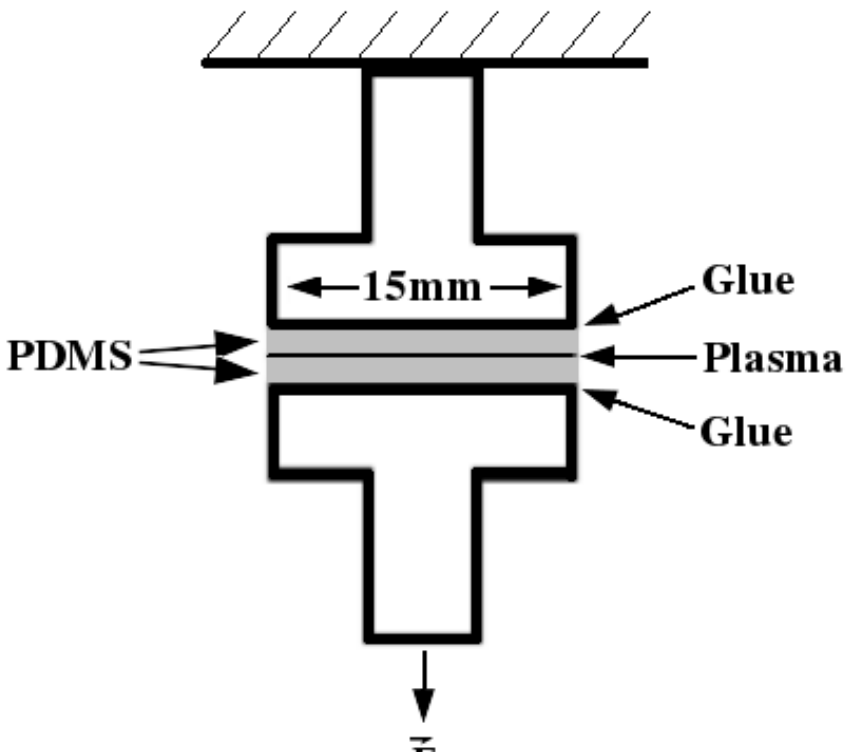

**Figure 3:** Scheme of the setup used to test the tensile strength supported between two PDMS samples bonded with DBD plasma.

## 4. Results and discussion

In order to evaluate the effect of plasma on the surface wettability, the WCA at the PDMS surfaces were measured before and immediately after the treatments. Figure 3(a) shows the WCA as function of the plasma treatment time for samples treated with He plasmas using the primary plasma jet (circles) and the transferred plasma jet (squares). The curves of recovery of WCA with aging time for PDMS samples treated for 90 s using both assemblies are shown in Fig. 4(b). The frequency of plasma pulses used to treat the samples was 60 Hz in all cases.

As can be seen in Fig. 4(a), WCA values tend to decrease monotonically with processing time in both cases, but the fastest WCA reduction was observed for the case of the transferred plasma jet, where the time needed to reach the minimum WCA value (~10°) was at least 30% lower than in the case of the primary discharge.

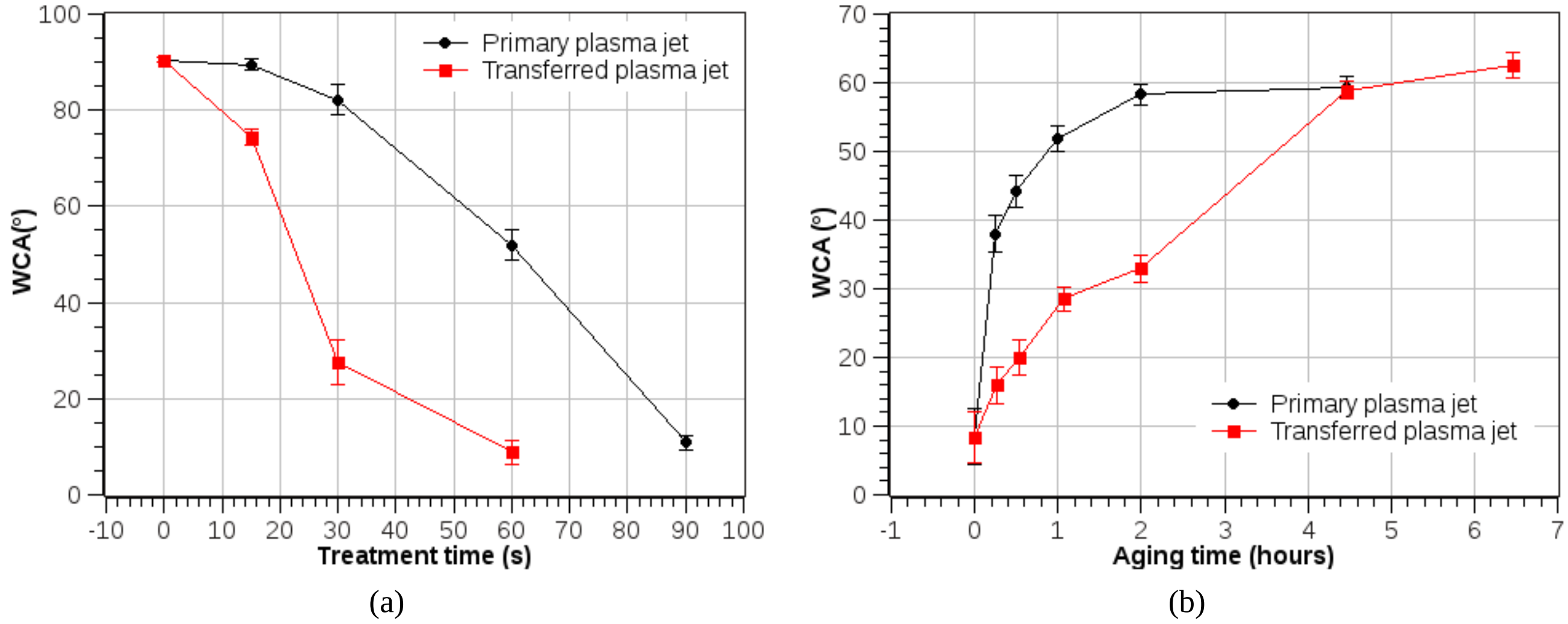

**Figure 4:** Comparison between treatments performed with primary plasma jet and transferred plasma jet for (a) WCA in function of plasma treatment time, and (b) WCA recovery in function of aging time.

In Fig. 4(b) we can see that the curves of WCA recovery with aging time are also quite different. When the primary plasma jet was used, the WCA values starts to rise fast from 10 to 40° in ~10-20 min, followed by slower increase to 60° in 2 hours. When the transferred plasma jet was used the WCA values rise more slowly in the first 2 hours and require 2 hours more to reach the value of 60°. In general, in the first two hours after the treatments, the WCA values for the samples treated with the transferred plasma jet are half of those for the samples treated with the primary plasma jet. Using the primary and the transferred plasma jets, we can say that the latter improved strongly the wettability of the PDMS surface in relation to the former comparing the reduction and recovery of WCA values.

The processes of reduction and recovery of WCA are likely to be related to chemical changes in the PDMS surface. The reduction is believed to be due to the surface functionalization by OH radicals [16,24,26]. This means that the transferred plasma jet outperforms the primary jet in efficiency of the PDMS surface functionalization The hydrophilic surface state is known to be unstable in air [42] and the WCA recovery begins immediately after the end of plasma treatment in both cases. However likely due to the larger density of OH radicals created in the PDMS surface with the transferred jet, the hydrophilic state remains active for a longer time.

Since during the recovery process the WCA does not reach its initial value of 90° during relatively long periods of time (Fig. 4b), it is likely that there is also a significant contribution of the reduction in the surface roughness to the decrease of WCA.

The surface morphology of PDMS samples were analyzed before and after the treatment with He plasmas of both assemblies using atomic force microscopy (AFM) analysis. Figures 5 (a) to (c) show AFM images of PDMS samples treated using the primary plasma jet; the samples were treated with 20 plasma pulses (a), during 30 s (~1800 plasma pulses) (b), and 5 min (~18000 plasma pulses) (c), respectively. Figures 5 (d) to (f) show AFM images of samples treated with the transferred plasma jet; 20 pulses in (d), 30 s (~1800 plasma pulses) in (e) and 5 min (~18000 plasma pulses) in (f). Figure 5(g) shows the sample that was not treated with plasma. Different samples were used for obtaining the AFM images. It can be seen that the surface roughness starts to change strongly (the surface becomes much flatter) when just a few plasma pulses are applied [Figs. 5(a) and 5(d)] and continues to change drastically with plasma treatment for longer processing times [Figs. 5(b)-(c) and 5(e)-(f)]. The data of the root mean square (RMS) surface roughness ($R_q$) obtained for all surfaces are presented in Table 1, where it can be seen that the minimum number of pulses reduces strongly the surface roughness in both assemblies used. It is important to observe that the reduction of surface roughness is much faster than the reduction of WCA, as can be seen from comparison with Fig. 4(a).

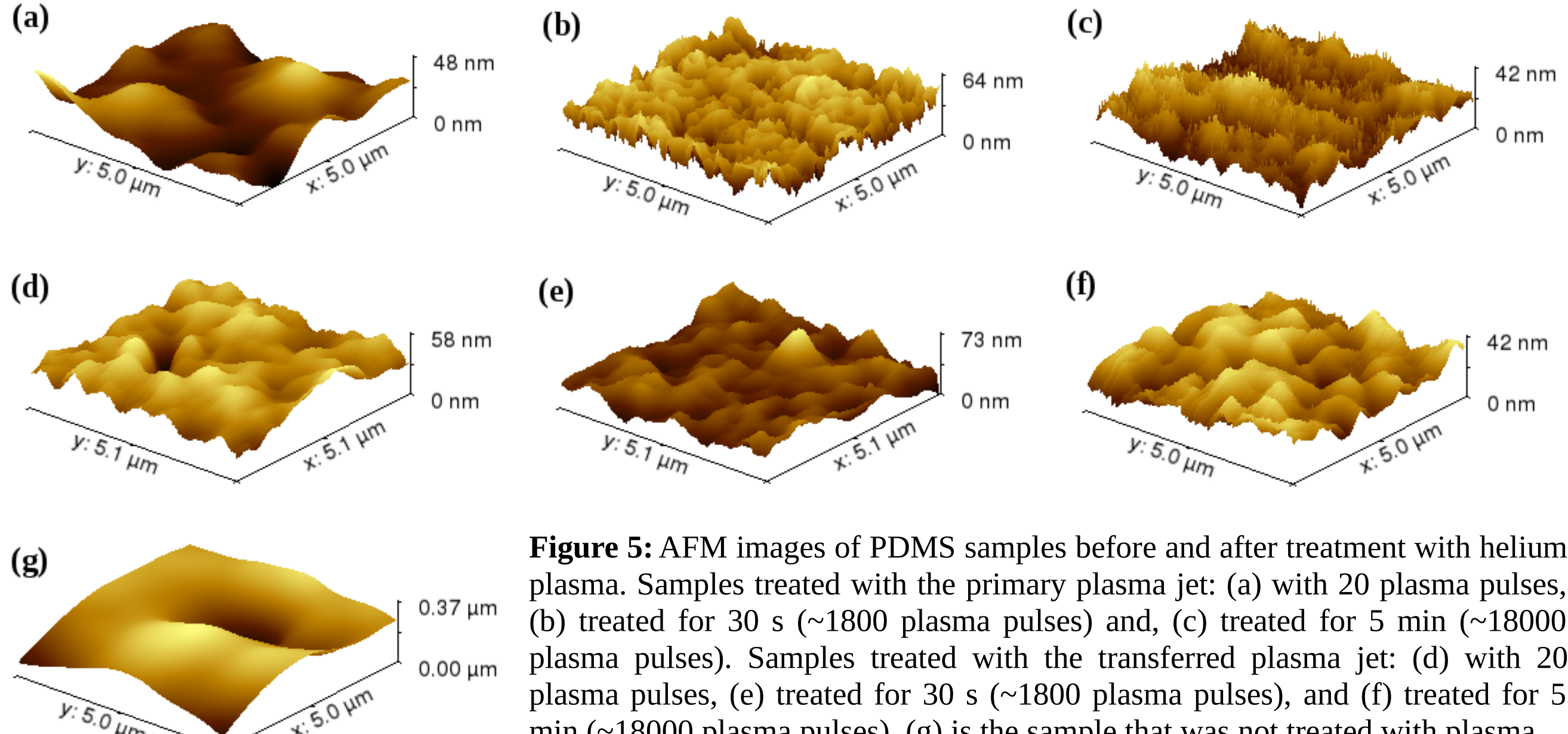

**Figure 5:** AFM images of PDMS samples before and after treatment with helium plasma. Samples treated with the primary plasma jet: (a) with 20 plasma pulses, (b) treated for 30 s (~1800 plasma pulses) and, (c) treated for 5 min (~18000 plasma pulses). Samples treated with the transferred plasma jet: (d) with 20 plasma pulses, (e) treated for 30 s (~1800 plasma pulses), and (f) treated for 5 min (~18000 plasma pulses). (g) is the sample that was not treated with plasma.

Moreover, although the surface roughness does not change so fast with time of treatment after the strong initial reduction, the surface morphology continues to change significantly as increasingly smaller scale features appear [compare Figs. 5(a)-4(c) and Figs. 5(d)-(f)], resulting in continuous increment of the surface area.

In general, when the surface roughness is considered, there are no significant differences between the PDMS surfaces treated with the primary or transferred plasma jets, in striking contrast to the evolution of WCA for both cases (Fig. 4).

**Table 1:** RMS roughness ($R_q$) values in function of the number of plasma pulses applied.

| **Number of pulses** | **$R_q$ (nm)** | |
|---|---|---|
| | Primary plasma | Transferred plasma |
| 0 | 40.95 ± 18.02 | |
| 20 | 7.3 ± 2.0 | 6.7 ± 1.9 |
| 1800 | 10.0 ± 1.3 | 6.8 ± 1.4 |
| 18000 | 5.14 ± 0.81 | 6.7 ± 1.4 |

Further, the measurements of tensile strength supported in adhesion between two similar PDMS samples were made following the procedure described in Sect. 3. For relatively long plasma treatment of the PDMS surfaces, the tensile strength supported in the adhesion was difficult to evaluate for both cases, because the adhesion failures were observed first for interfaces between PDMS and metal screws, since the glue used was found to support a tensile strength not higher than ~2.0 kgf/cm$^2$. In other words, the tensile strength required for samples detachment was greater than 2.0 kgf/cm$^2$. In order to reveal the differences in the treatment with the primary plasma jet and the transferred plasma jet, additional tests were performed by applying only 20 plasma pulses to each PDMS sample, in this case the adhesion failures were observed first at the PDMS-PDMS interfaces.

Figure 6 shows the results of measurements of the tensile strength supported in adhesion between two PDMS samples treated using the primary and transferred helium plasma jets (denoted as He and He-JT, in the figure). Fig. 6 also shows the tensile strength supported in the case where the surface treatment was made with the primary plasma jet using nitrogen gas to create the plasma ($N_2$ in the figure). The values of the tensile strength shown in Fig. 6 were obtained as an average for 5 samples for each case. Furthermore, Fig. 6 shows, for each type of plasma, the vibrational temperatures calculated using the plasma spectra. A clear correlation of the tensile strength results with the vibrational temperature can be also seen. The values obtained for the vibrational temperatures were 2300, 2800 and 3300 K, for $N_2$, He and He-JT, respectively, with estimated errors of 50 K in all cases.

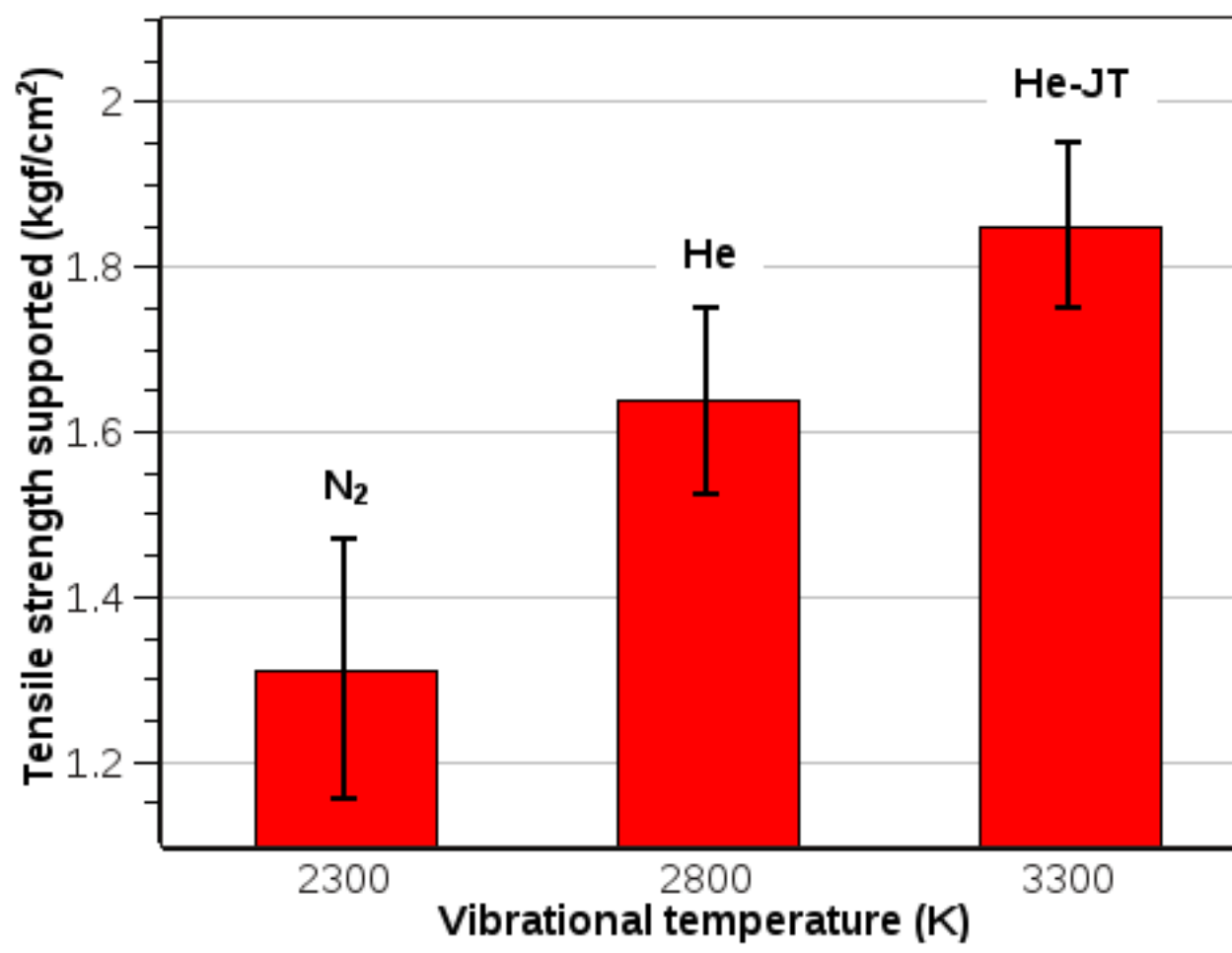

**Figure 6:**Tensile strength supported by PDMS-PDMS adhesion for different plasma jets and different vibrational temperatures of them.

As we can see in Fig. 6, the best result (maximum tensile strength) for adhesion was obtained for the treatment made with the transferred He plasma jet (1.85 ± 0.10 kgf/cm$^2$), followed by the primary He plasma jet (1.64 ± 0.11 kgf/cm$^2$) and finally by $N_2$ primary plasma jet (1.31 ± 0.16 kgf/cm$^2$).

The rotational temperatures of the plasma jets were measured also, and we obtained similar values (400 ± 50 K) for both the primary and transferred He plasma jets. Thus the rotational temperature ($T_{rot}$), that is known to be approximately equal to the gas temperature ($T_{gas}$) [43-45], seemingly does not affect the adhesion properties.

In our previous work [13], we measured the power delivered to the primary plasma jet and we found that it was equal to 7.8 ± 0.6 mW, using He gas, and 10.6 ± 1.1 mW using $N_2$ gas, when operating at six plasma pulses per second. That power calculations were made using a well known method, measuring simultaneously, in one plasma pulse, the voltage applied on the electrode and the current across a shunt resistor [46]. When working with the transferred plasma jet we were not able to measure the current across the shunt resistor because the signal to noise ratio was very low in this case, which means that the current is much lower in this case making the power estimates unfeasible. Kostov *et al* [28] demonstrated that the voltage at the end tip of the Cu wire is much lower than the voltage applied to the electrode, and that the voltage decreases when the length of the wire is increased. Then, considering the reduction of the current across the shunt resistor and the decrease of the voltage in the Cu wire, we can expect that the power delivered to the transferred plasma jet is considerably smaller than for the case of a primary plasma jet. Further studies are being conducted in order to determine correctly the power delivered to the transferred plasma jet in our device.

The emission spectra obtained for the primary and transferred plasma jets are shown in Fig. 7.

The spectra in Fig. 7 are accumulations of 1000 plasma pulses acquired using a 150 lines/mm grating. Regarding the species found in the emission spectra, we can see that there are no differences between the primary and transferred plasma jets. In both cases there are many line emissions of molecular $N_2$ ($N_2$ I), two line emissions from molecular $N_2^+$ ion ($N_2$ II), and emissions of OH radical are also present, with relatively weaker intensities from molecular ions in the case of transferred jet. Then, the changes in the surface chemistry of PDMS after plasma treatments are not due to chemical changes in the plasma but to changes in its physical properties.

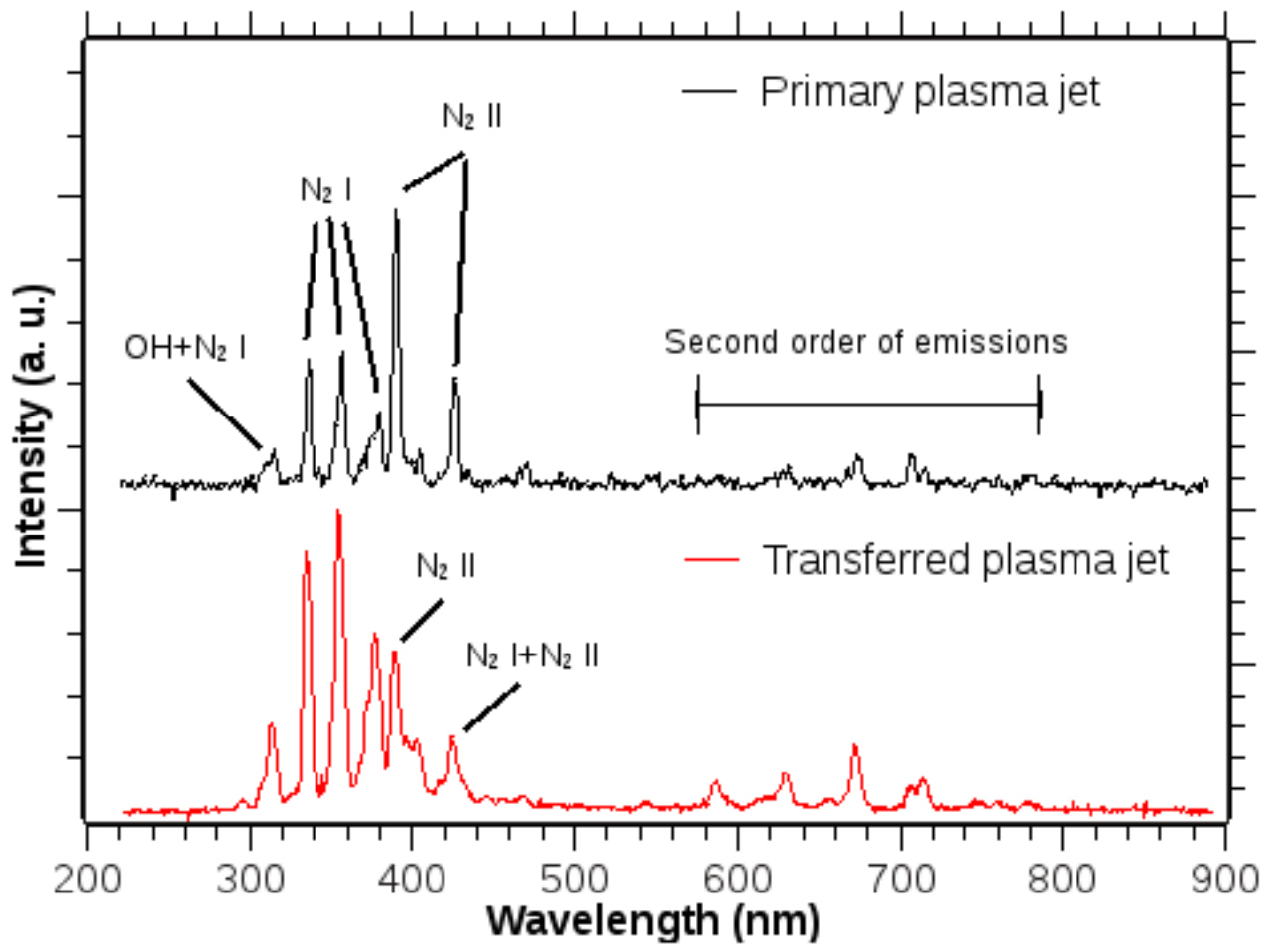

**Figure 7:**Emission spectra of primary and transferred plasma jets.

Considering the balance of vibrational energy in the plasmas, the conservation equation for vibrational energy of excited nitrogen molecules $E_{vib}(T_{vib})$ can be written as [47, 48]:

$$\frac{dE_{vib}(T_{vib})}{dt} = \alpha_v U_E - \frac{E_{vib}(T_{vib}) - E_0(T_{gas})}{\tau_{VT}(T_{gas})} \qquad (1)$$

where $E_0(T_{gas})$ is the equilibrium vibrational energy, $\alpha_v$ is the rate in which the electromagnetic field energy is transferred into the energy of vibrational states, $U_E$ is the electromagnetic field energy, and $\tau_{VT}(T_{gas})$ is the vibrational-translational relaxation time.

If we assume that the deviation from the thermal nonequilibrium is small, for the nonequilibrium vibrational energy we can write [47]:

$$E_{vib}(T_{vib}) = N \frac{h\nu_0}{\exp(h\nu_0 / T_{vib}) - 1} \qquad (2)$$

where N is the gas number density and $h\nu_0$ is the energy of the nitrogen vibrational quantum. From Eq. 2 we can see that the vibrational energy increases with the vibrational temperature.

In order to analyze the reason for the increase of the vibrational temperature when the transferred plasma jet is used, we are not interested in the time variation of the vibrational energy with time and make and we can assume $dE_{vib}/dt = 0$ in Eq. 1. Since there is no significant differences in the gas temperature between the primary and transferred plasma jets, we can analyze just the case:

$$\alpha_v U_E \propto E_{vib} / \tau_{VT} \quad (3)$$

Since N does not depend explicitly on the vibrational temperature, we can rewrite (3) as:

$$\alpha_v \frac{U_E}{N} \propto \frac{E_{vib}}{N \tau_{VT}} \propto (\exp(h \nu_0 / T_{vib}) - 1)^{-1} \quad (4)$$

The voltage in the transferred plasma jet ($V_T$) is lower than that in the primary plasma jet ($V_P$) as discussed above, which causes a reduction in the electric field energy ($U_E/N$). Then, we can infer that the reason for the increase in the vibrational temperature in the transferred plasma jet is that the $\alpha_v$ term increases by a factor larger than the $U_E$ /N ratio decreases. The increase in the rate $\alpha_v$ implies the increase in the average time that molecules remain in the higher energy states. In other words, a larger population of molecules must be in the higher vibrational states. Possible evidence of this is presented in Figure 8 that shows $N_2$ I emission lines (second positive, C $^3\Pi_u$ – B $^3\Pi_g$ transitions) in the spectral range from 365 to 385 nm from both plasmas. The comparison between experimental and simulated curves was used to estimate the rotational and vibrational temperatures for nitrogen molecules. However, Fig. 8 also shows the population distribution over the vibrational energy levels. The energy of the vibrational levels increases from peak 1 to peak 4. As we can see in Fig. 8, the population of higher energy levels (peaks 2 to 4) is systematically higher in comparison with the population of the lowest energy level (peak 1) when we compare the primary plasma jet with the transferred one ($I_2/I_1$ are equal to ~0.65 and ~0.55, $I_3/I_1$ are equal to ~0.20 and ~0.12, etc.). This can be considered as an indication that the $\alpha_v$ term (the rate in which the electromagnetic field energy is transferred into the energy of vibrational states) is significantly higher for the transfered jet.

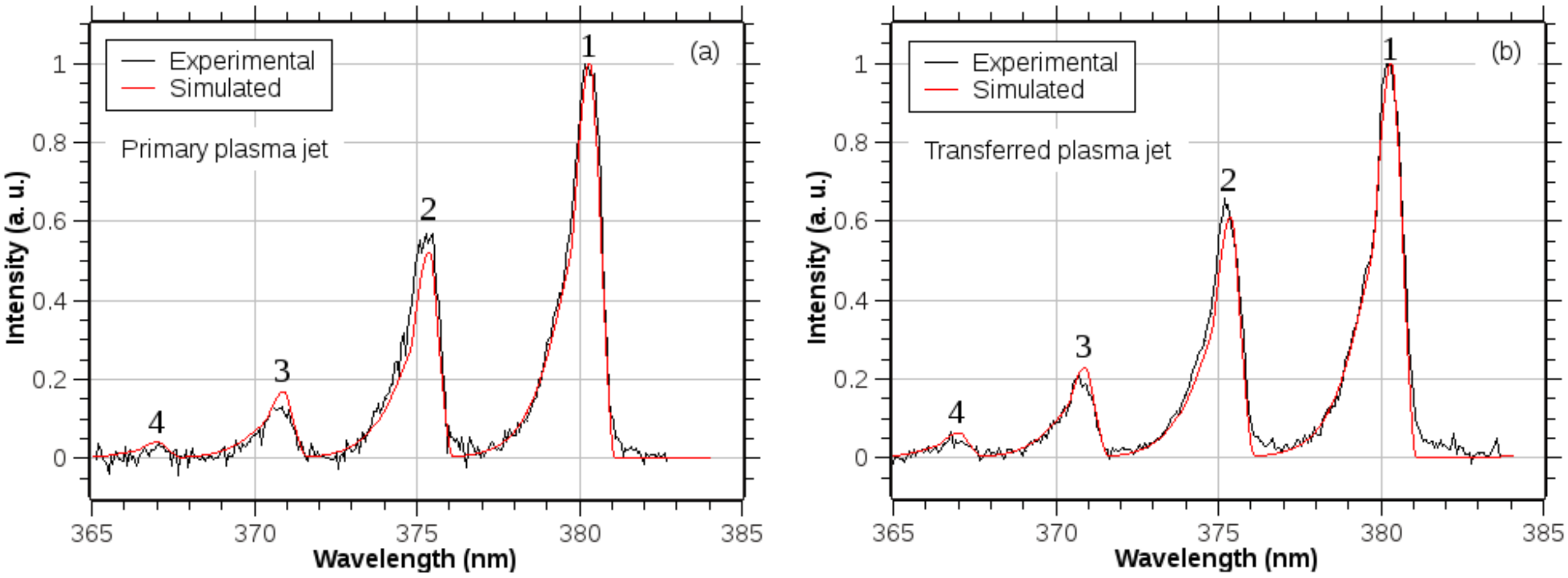

**Figure 8:** Plots of experimental and simulated emission spectra used to estimate rotational and vibrational temperatures for primary (a) and transferred (b) plasma jets.

The reasons for the observed differences between the two assemblies that may be related to the increase in $T_{vib}$ and in $\alpha_v$ parameters are not clear at the moment and this will the subject of future studies.

## 5. Conclusions

The main conclusion of this work is that when using transferred plasma jet instead of the primary jet discharge, a higher vibrational temperature was obtained, and this results in significant improvement of the polymer surface treatment and its adhesion properties. The other parameters that might be considered are plasma power, flux of reactive species, and the other temperatures associated with the plasma (rotational, translational and electron temperatures). We saw that the plasma power of the transferred plasma jet is lower than the primary plasma jet. The spectrum obtained in both plasmas showed that there are no differences between the species found in the plasmas. It is unlikely that the electron temperature has some effect on surface modification. The rotational temperature remained unchanged when changing from one to another plasma, and since we have $T_{rot} \approx T_{gas}$, the translational temperature also remained unchanged.

The results in the WCA reduction and recovery after processing show that the use of transferred plasma jet is more advantageous for the treatment of PDMS surfaces because it reduces the time required to reduce the WCA to its minimum value and retards the recovery of WCA.

There is a clear correlation between the tensile strength supported in adhesion and the vibrational temperature of the plasma jet used to treat the PDMS surfaces. The adherence between surfaces and the tensile strength seem to be proportional to the vibrational temperature.

In the surface treatment with plasma, the gas vibrational temperature is unlikely to induce significant changes in surface roughness, but, according to the WCA curves, it plays an important role in the surface activation and creation of functional groups that are responsible for the improved adhesion.

The most likely reason for the increase in the vibrational temperature when the transferred plasma jet is used is an increase in the rate in which the electromagnetic field energy is transferred into the energy of vibrational states.

## Acknowledgments

This work was supported by CAPES and CNPq.